\documentclass[preprint,prb,amsmath,amssymb,superscriptaddress]{revtex4}

\usepackage{graphicx}
\usepackage{dcolumn}
\usepackage{bm}
\usepackage[sort&compress]{natbib}

\begin{document}

\title{Strong coupling between quantum dot exciton spin states and a photonic crystal cavity }

\author{Hyochul Kim} \email{hckim@umd.edu}
\affiliation{Department of Electrical and Computer Engineering, IREAP, University of Maryland, College Park,
Maryland 20742, USA and Joint Quantum Institute, University of Maryland, College Park, Maryland 20742, USA}

\author{Thomas C. Shen}
\affiliation{Department of Electrical and Computer Engineering, IREAP, University of Maryland, College Park,
Maryland 20742, USA and Joint Quantum Institute, University of Maryland, College Park, Maryland 20742, USA}

\author{Deepak Sridharan}
\affiliation{Department of Electrical and Computer Engineering, IREAP, University of Maryland, College Park,
Maryland 20742, USA and Joint Quantum Institute, University of Maryland, College Park, Maryland 20742, USA}

\author{Glenn S. Solomon}
\affiliation{Joint Quantum Institute, National Institute of Standards and Technology, and University of Maryland, Gaithersburg, Maryland 20899, USA}

\author{Edo Waks}
\affiliation{Department of Electrical and Computer Engineering, IREAP, University of Maryland, College Park,
Maryland 20742, USA and Joint Quantum Institute, University of Maryland, College Park, Maryland 20742, USA}

\date{\today}

\begin{abstract}
We apply magnetic fields of up to 7~T to an indium arsenide (InAs) quantum dot (QD) strongly coupled to a photonic crystal cavity.  The field lifts the degeneracy of QD exciton spin states, and tune their emission energy by a combination of diamagnetic and Zeeman energy shifts.  We use magnetic field tuning to shift the energies of the two spin states to be selectively on resonance with the cavity. Strong coupling between the cavity and both spin states is observed.  Magnetic field tuning enables energy shifts as large as 0.83~meV without significant degradation of the QD-cavity coupling strength.
\end{abstract}

\maketitle

Semiconductor QDs coupled to optical microcavities provide a promising physical platform for studying cavity quantum electrodynamics (cQED).  These systems provide large matter-light interaction strengths owing to the large oscillator strengths of QDs and the high quality factors and small mode volumes of various microcavity designs, enabling seminal studies of cQED in the strong coupling regime\cite{ReithmaierNat04,YoshieNat04,PeterPRL05}.  This regime opens up the possibility for controlling and modifying cavity reflectivity via QD interactions\cite{WaksPRL06,RakherPRL09,EnglundNat07} and exhibits nonlinear optical effects near the single photon level\cite{FushmanSci08}.  Strong coupling also provides a promising avenue for generating coherent interactions between QDs and optical fields, forming a basic building block for all-optical solid-state quantum networking and quantum computation\cite{ImamogluPRL99,SridharanPRA08}.

The application of magnetic fields to QDs that are coupled to optical microcavities provides access to additional degrees of freedom based on QD spin or photon angular momentum.  The magnetic fields serve to lift the degeneracy of these states of the QD, providing a more complex level structure\cite{BayerPRB02} with important implications for coherent light-matter interaction and quantum information processing. These states have already been used to implement coherent population trapping\cite{XuNPhys08} and quantum control of a single QD spin\cite{PressNat08} in bare QDs.  Incorporation of optical microcavities is expected to enhance these effects and enable additional applications such as single photon generation with improved indistinguishability\cite{KirazPRA04}, and entanglement between QDs and optical fields\cite{WaksPRL06,SridharanPRA08}. In addition, magnetic fields provide a practically useful tuning mechanism to control interactions between QD emission and a microcavity.  In contrast to methods such as temperature tuning\cite{YoshieNat04,FaraonAPL07}, local gas deposition\cite{MosorAPL05,StraufAPL06}, DC stark shifts\cite{LauchtNJP09,KimAPL09}, and photochromic films\cite{SridharanAPL10}, magnetic fields enable both red and blue shifting depending on the spin state of the QD.

Recently, magnetic field tuning was used to tune the emission of an elongated indium gallium arsenide (InGaAs) QD into resonance with micropillar cavity systems\cite{ReitzensteinPRL09}.  In these systems, tuning was achieved by diamagnetic shifting rather than Zeeman spin splitting.  At large magnetic fields (above 3~T), magnetic confinement of electrons and holes was found to degrade the cavity-QD coupling strength due to reduction of the QD dipole moment\cite{ReitzensteinPRL09}.  In this letter, we demonstrate strong coupling between exciton spin states of an InAs QD coupled to a photonic crystal cavity in the strong coupling regime. By using Zeeman split exciton lines of InAs QDs, we were able to tune the QD emission frequency over 0.83~meV with applied magnetic fields as high as 7~T. In contrast to elongated InGaAs QDs, we show that the coupling between InAs QD spin states and the cavity mode is largely independent of magnetic field strength, allowing large tuning ranges without significant reduction in the cavity-QD coupling strength.  Ultimately, this system provides a three level 'V' system inside the cavity that may have important applications for quantum information and basic studies of quantum coherence.

Measurements were performed on self assembled InAs QDs coupled to gallium arsenide (GaAs) photonic crystal cavities. The initial sample consisted of a 160~nm thick GaAs layer grown on top of a 1~$\mu$m aluminum gallium arsenide (Al$_{0.78}$Ga$_{0.22}$As) sacrificial layer. InAs QDs were grown in the middle of the GaAs layer.  Photonic crystal cavities were patterned into the structure by electron beam lithography and inductively coupled plasma etching, followed by wet etch removal of the sacrificial layer. Photonic crystal cavities used in this experiment were based on a three hole defect (L3) cavity with fine tuning of the six holes near the cavity edge\cite{AkahaneOptExp05}.  A scanning electron microscope image of a typical cavity structure is shown in the inset to Figure 1(a). The high power photoluminescence (PL) spectrum from the cavity device used in these experiments is shown in Figure 1(a).  From the PL measurement, the cavity Q is determined to be 9,000.

The sample was mounted in continuous flow liquid helium cryostat and cooled down to a temperature of $10-50$~K.  The cold finger of the cryostat was surrounded by a superconducting magnet that can apply magnetic fields of up to 7~T.  The direction of the magnetic field was parallel to the sample growth direction (Faraday configuration). The sample was excited by a Ti:Sapphire laser tuned to 865~nm, and the emission was collected by a confocal microscope using an objective lens (numerical aperture~$= 0.7$). The spectrum of the collected signal was measured by a grating spectrometer with a resolution of 0.02~nm.

We initially measured the PL from a bare QD in the absence of any cavity as a function of a magnetic field as shown in Figure 1(b). When a magnetic field was applied, the QD emission split into two branches corresponding to exciton states with angular momentum of -1 for the higher energy branch and +1 for the low energy branch. The energy shift of the two exciton states is caused by both the Zeeman energy splitting $\Delta \textrm{E}_{Zeeman}=\gamma_1 \textrm{B}$ and diamagnetic shift $\Delta \textrm{E}_{dia}=\pm \gamma_2 \textrm{B}^2$ where $\gamma_1=(g_e-g_h) \mu_\textrm{B}/2$ and $\gamma_2$ is the diamagnetic coefficient\cite{BayerPRB02,WalckPRB98}. Here $g_e$ and $g_h$ are electron and hole $g$ factors and $\mu_\textrm{B}$ is the Bohr magneton. The fitting to the QD PL curve gives $g_e-g_h=2.9$ and $\gamma_2=6~\mu$eV/T$^2$, which are consistent with previous measurements\cite{BayerPRB02,ReitzensteinPRL09}. At 7~T, we achieve a red shift of 0.21~nm (0.3~meV) for the +1 exciton state and a blue shift of 0.58~nm (0.83~meV) for the -1 exciton state. These tuning ranges are sufficiently large to selectively bring only one of the excitonic states onto resonance with the cavity. It is worth noting that neutral excitons and charged excitons show identical Zeeman behavior as the splitting is given by the difference between electron and hole $g$ factors in both cases\cite{KrizhanovskiiPRB05}.

Figure 2 shows strong coupling of the two Zeeman split branches of QD emission to the cavity mode. The sample temperature is initially set to 34~K where the QD emission wavelength is blue shifted from the cavity mode by 0.12~nm with no magnetic field. The spin +1 exciton emission is then shifted onto cavity resonance by application of a magnetic field, as shown in Figure 2(a). At around 2.1~T the exciton emission becomes resonant with the cavity mode.  The two emission lines exhibit a clear anti-crossing indicating that the exciton state is strongly coupled to the cavity mode. We next change the temperature of the sample to 41~K to shift the QD emission so that it is red detuned to the cavity by 0.22~nm.  The magnetic field was once again applied to blue shift the -1 exciton state through the cavity mode, as shown in Figure 2(b).  Again, strong coupling between the cavity mode and spin -1 exciton state of the same QD was observed at about 2.7~T, showing that both exciton branches can strongly couple to the cavity mode.

We next investigate the effect of the magnetic field strength on the coupling between the exciton states and the cavity.  These measurements were performed by temperature tuning the QD in the presence of a fixed magnetic field. The temperature was varied over a sufficiently wide range so that both excitonic states were swept across the cavity resonance. Figure 3(a) shows the temperature tuning spectra at 0~T. In this case there is no splitting between the two spin states of the exciton so only a single anti-crossing is observed.  The QD line shifts on resonance with the cavity mode at around 33~K. On resonance, the cavity mode and QD emission line are strongly coupled, showing minimum energy separation of $\Delta \textrm{E}=123~\mu$eV.   This number can be used to extract the cavity-QD coupling strength $g$ using the equation
\begin{equation}
\Delta \textrm{E}=2 \sqrt{g^2-\frac{(\gamma_c-\gamma_x)^2}{16}}
\end{equation}
where  $\gamma_c$ and $\gamma_x$ are the linewidths of the cavity mode and exciton respectively. The cavity linewidth is determined to be $\gamma_c= 150~\mu$eV from the quality factor ($Q=9,000$) while the QD exciton linewidth of $\gamma_x= 1~\mu$eV is taken from the literature\cite{MullerPRL09}. Using these numbers, the vacuum Rabi frequency at 0~T is determined to be $g = 72~\mu$eV.  At 1~T, the two exciton states are split by 0.11~nm separation (Figure 3(a)). Both the +1 and -1 exciton states show clear strong coupling to the cavity mode at 32~K and 35~K, respectively. The minimum energy separations are $\Delta \textrm{E}=102~\mu$eV for the +1 spin state and $\Delta \textrm{E}=94~\mu$eV for the -1 spin state, corresponding to the Rabi frequency of $63~\mu$eV and $60~\mu$eV respectively.

Temperature tuning measurements were taken for magnetic fields ranging from 1~T to 7~T.  For each magnetic field value, the Rabi frequency was calculated from the energy separation of the polariton peaks at the strong coupling point.  The results of this measurement are shown in Figure 3(b). Red circles indicate the calculated Rabi frequency for the +1 exciton, while blue squares indicate the value for the -1 exciton.  The black triangle indicates the Rabi frequency of the exciton state at 0~T.  When no magnetic field is present, anisotropic hyperfine splitting due to strain is expected to split the neutral exciton emission into linear polarized components that lie roughly parallel to the in-plane GaAs crystal direction ($110$ and $1\overline{1}0$).  The photonic crystal cavities were fabricated such that their preferred polarization lies parallel to the in-plane crystal axis as well. If the orientation of the QD dipole is perfectly aligned with the cavity polarization axis, one would expect that at higher magnetic field the Zeeman splitting would dominate the anisotropic splitting resulting in a $g'=g_0/\sqrt{2}$, where $g'$ is the Rabi frequency at high fields and $g_0$ is the Rabi frequency in the absence of a magnetic field.  The $\sqrt{2}$ reduction in the coupling strength is due to a change of selection rules from linearly polarized to circularly polarized emission. From Figure 3(b) we see that a reduction of 11\% for the +1 exciton and 17\% for the -1 exciton is observed (when averaging over the different magnetic fields).  This reduction is smaller than what is expected from a perfectly oriented dipole moment, suggesting that the initial dipole orientation of the QD at 0~T was not exactly parallel to the cavity axis\cite{MullerPRL09}.  Figure 3(b) also shows that between 1~T and 7~T, the coupling strength $g$ is not strongly dependent on the magnetic field strength, which is consistent with the lateral extension of the electron and hole wave functions of a single InAs QD being smaller than the magnetic confinement length at 7~T\cite{ReitzensteinPRL09,HKim}.  This result shows that for InAs QDs coupled to photonic crystal cavities we can use large magnetic fields to tune excitonic emission without significantly sacrificing the coupling strength.

In summary, we demonstrated that the degenerate QD exciton states of an InAs QD can be lifted by applying a magnetic field and that each state can be strongly coupled to the photonic crystal cavity mode. The small mode volume of photonic crystal cavities allows strong coupling between a single QD and a cavity mode even at high magnetic field (7~T). In addition to serving as a method for tuning QD emission relative to the cavity resonance, application of magnetic fields may open the possibility for exploring the interactions between a cavity and multi-level atomic system.  For instance, neutral or charged excitons in magnetic field (Faraday or Voigt configurations) create effective '$\Lambda$' or 'V' multi level states. Such three level atomic configurations can be exploited to achieve optically controlled cavity reflectivity.  In addition, magnetic field tuning can serve as an effective method to bring two QDs having a small emission frequency difference onto resonance with a single cavity mode by using the red shifted branch of one QD and the blue shifted branch of the other QD.

The authors would like to acknowledge support from the ARO MURI on Hybrid quantum interactions (grant number W911NF09104), the physics frontier center at the joint quantum institute, and the ONR applied electromagnetic center.  E. Waks would like to acknowledge support from an NSF CAREER award (grant number ECCS. 0846494).

\pagebreak

\begin{figure}
\centering\includegraphics[width=8cm]{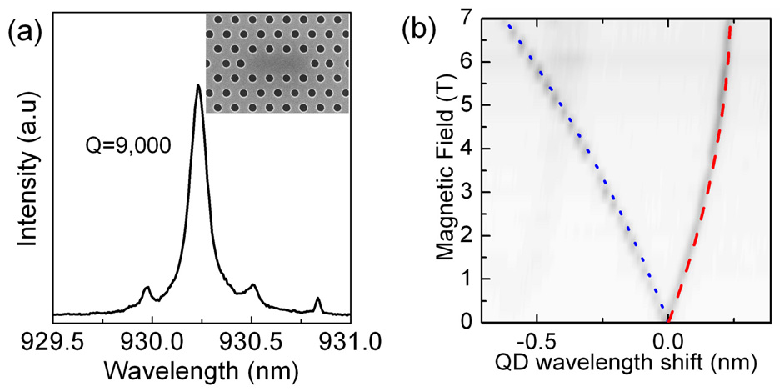}
\caption{(a) Photoluminescence spectrum of photonic crystal cavity showing a Q of 9,000. A scanning electron microscope image of a typical fabricated L3 cavity is shown in the inset. (b) Single QD emission is split into spin +1 and -1 exciton branches by an applied magnetic field in the absence of any cavity. The blue dotted line and red dashed line show calculated fits.}
\end{figure}

\begin{figure}[b]
\centering\includegraphics[width=8cm]{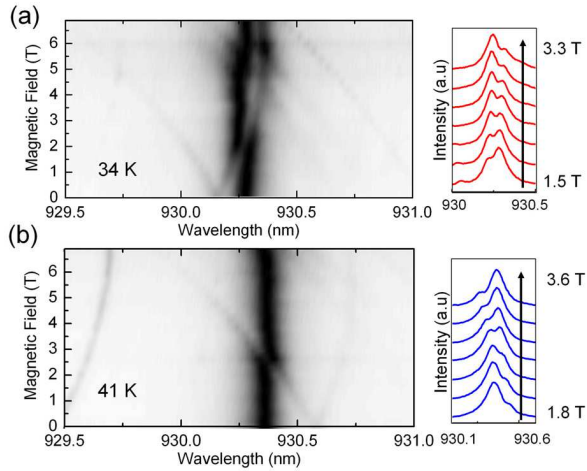}
\caption{Strong coupling of exciton spin states to a cavity mode.  (a) Magnetic field tuning when sample is held at 34~K (QD is blue shifted from cavity) and +1 exciton is shifted onto resonance. (b) Magnetic field tuning when sample is held at 41~K (QD is red shifted from cavity) and -1 exciton is blue shifted onto resonance. Right panels show measured spectra from 1.5~T to 3.3~T (0.3~T steps) for +1 exciton and 1.8~T to 3.6~T for -1 exciton.}
\end{figure}

\begin{figure}
\centering\includegraphics[width=8cm]{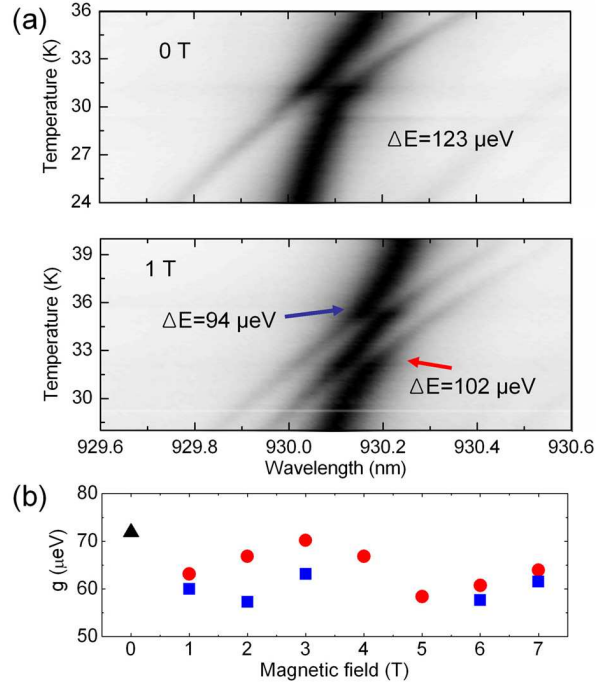}
\caption{(a) Spectra of coupled QD-cavity mode with temperature scanning at 0~T (top) and 1~T (bottom): The QD exciton emission red-shifts and crosses the cavity mode as temperature increases. The minimum energy separation $\Delta \textrm{E}$ is indicated at the anti-crossing point. (b) Measured Rabi frequency based on the minimum energy separation: coupling strength at 0~T is indicated by a black triangle and +1 (-1) exciton coupling constants are indicated by red circles (blue rectangles).}
\end{figure}


\begin{thebibliography}{25}
\expandafter\ifx\csname natexlab\endcsname\relax\def\natexlab#1{#1}\fi
\expandafter\ifx\csname bibnamefont\endcsname\relax
  \def\bibnamefont#1{#1}\fi
\expandafter\ifx\csname bibfnamefont\endcsname\relax
  \def\bibfnamefont#1{#1}\fi
\expandafter\ifx\csname citenamefont\endcsname\relax
  \def\citenamefont#1{#1}\fi
\expandafter\ifx\csname url\endcsname\relax
  \def\url#1{\texttt{#1}}\fi
\expandafter\ifx\csname urlprefix\endcsname\relax\def\urlprefix{URL }\fi
\providecommand{\bibinfo}[2]{#2}
\providecommand{\eprint}[2][]{\url{#2}}

\bibitem[{\citenamefont{{Reithmaier} et~al.}(2004)\citenamefont{{Reithmaier},
  {S{\c e}k}, {L{\"o}ffler}, {Hofmann}, {Kuhn}, {Reitzenstein}, {Keldysh},
  {Kulakovskii}, {Reinecke}, and {Forchel}}}]{ReithmaierNat04}
\bibinfo{author}{\bibfnamefont{J.~P.} \bibnamefont{{Reithmaier}}},
  \bibinfo{author}{\bibfnamefont{G.}~\bibnamefont{{S{\c e}k}}},
  \bibinfo{author}{\bibfnamefont{A.}~\bibnamefont{{L{\"o}ffler}}},
  \bibinfo{author}{\bibfnamefont{C.}~\bibnamefont{{Hofmann}}},
  \bibinfo{author}{\bibfnamefont{S.}~\bibnamefont{{Kuhn}}},
  \bibinfo{author}{\bibfnamefont{S.}~\bibnamefont{{Reitzenstein}}},
  \bibinfo{author}{\bibfnamefont{L.~V.} \bibnamefont{{Keldysh}}},
  \bibinfo{author}{\bibfnamefont{V.~D.} \bibnamefont{{Kulakovskii}}},
  \bibinfo{author}{\bibfnamefont{T.~L.} \bibnamefont{{Reinecke}}},
  \bibnamefont{and}
  \bibinfo{author}{\bibfnamefont{A.}~\bibnamefont{{Forchel}}},
  \bibinfo{journal}{Nature (London)} \textbf{\bibinfo{volume}{432}},
  \bibinfo{pages}{197} (\bibinfo{year}{2004}).

\bibitem[{\citenamefont{{Yoshie} et~al.}(2004)\citenamefont{{Yoshie},
  {Scherer}, {Hendrickson}, {Khitrova}, {Gibbs}, {Rupper}, {Ell}, {Shchekin},
  and {Deppe}}}]{YoshieNat04}
\bibinfo{author}{\bibfnamefont{T.}~\bibnamefont{{Yoshie}}},
  \bibinfo{author}{\bibfnamefont{A.}~\bibnamefont{{Scherer}}},
  \bibinfo{author}{\bibfnamefont{J.}~\bibnamefont{{Hendrickson}}},
  \bibinfo{author}{\bibfnamefont{G.}~\bibnamefont{{Khitrova}}},
  \bibinfo{author}{\bibfnamefont{H.~M.} \bibnamefont{{Gibbs}}},
  \bibinfo{author}{\bibfnamefont{G.}~\bibnamefont{{Rupper}}},
  \bibinfo{author}{\bibfnamefont{C.}~\bibnamefont{{Ell}}},
  \bibinfo{author}{\bibfnamefont{O.~B.} \bibnamefont{{Shchekin}}},
  \bibnamefont{and} \bibinfo{author}{\bibfnamefont{D.~G.}
  \bibnamefont{{Deppe}}}, \bibinfo{journal}{Nature (London)}
  \textbf{\bibinfo{volume}{432}}, \bibinfo{pages}{200} (\bibinfo{year}{2004}).

\bibitem[{\citenamefont{{Peter} et~al.}(2005)\citenamefont{{Peter},
  {Senellart}, {Martrou}, {Lema{\^i}tre}, {Hours}, {G{\'e}rard}, and
  {Bloch}}}]{PeterPRL05}
\bibinfo{author}{\bibfnamefont{E.}~\bibnamefont{{Peter}}},
  \bibinfo{author}{\bibfnamefont{P.}~\bibnamefont{{Senellart}}},
  \bibinfo{author}{\bibfnamefont{D.}~\bibnamefont{{Martrou}}},
  \bibinfo{author}{\bibfnamefont{A.}~\bibnamefont{{Lema{\^i}tre}}},
  \bibinfo{author}{\bibfnamefont{J.}~\bibnamefont{{Hours}}},
  \bibinfo{author}{\bibfnamefont{J.~M.} \bibnamefont{{G{\'e}rard}}},
  \bibnamefont{and} \bibinfo{author}{\bibfnamefont{J.}~\bibnamefont{{Bloch}}},
  \bibinfo{journal}{\prl} \textbf{\bibinfo{volume}{95}},
  \bibinfo{pages}{067401} (\bibinfo{year}{2005}).

\bibitem[{\citenamefont{Waks and Vuckovic}(2006)}]{WaksPRL06}
\bibinfo{author}{\bibfnamefont{E.}~\bibnamefont{Waks}} \bibnamefont{and}
  \bibinfo{author}{\bibfnamefont{J.}~\bibnamefont{Vuckovic}},
  \bibinfo{journal}{\prl} \textbf{\bibinfo{volume}{96}}, \bibinfo{eid}{153601}
  (\bibinfo{year}{2006}).

\bibitem[{\citenamefont{Rakher et~al.}({2009})\citenamefont{Rakher, Stoltz,
  Coldren, Petroff, and Bouwmeester}}]{RakherPRL09}
\bibinfo{author}{\bibfnamefont{M.~T.} \bibnamefont{Rakher}},
  \bibinfo{author}{\bibfnamefont{N.~G.} \bibnamefont{Stoltz}},
  \bibinfo{author}{\bibfnamefont{L.~A.} \bibnamefont{Coldren}},
  \bibinfo{author}{\bibfnamefont{P.~M.} \bibnamefont{Petroff}},
  \bibnamefont{and}
  \bibinfo{author}{\bibfnamefont{D.}~\bibnamefont{Bouwmeester}},
  \bibinfo{journal}{\prl} \textbf{\bibinfo{volume}{{102}}},
  \bibinfo{pages}{097403} (\bibinfo{year}{{2009}}).

\bibitem[{\citenamefont{Englund et~al.}(2007)\citenamefont{Englund, Faraon,
  Fushman, Stoltz, Petroff, and Vu\v{c}kovi\'{c}}}]{EnglundNat07}
\bibinfo{author}{\bibfnamefont{D.}~\bibnamefont{Englund}},
  \bibinfo{author}{\bibfnamefont{A.}~\bibnamefont{Faraon}},
  \bibinfo{author}{\bibfnamefont{I.}~\bibnamefont{Fushman}},
  \bibinfo{author}{\bibfnamefont{N.}~\bibnamefont{Stoltz}},
  \bibinfo{author}{\bibfnamefont{P.}~\bibnamefont{Petroff}}, \bibnamefont{and}
  \bibinfo{author}{\bibfnamefont{J.}~\bibnamefont{Vu\v{c}kovi\'{c}}},
  \bibinfo{journal}{Nature (London)} \textbf{\bibinfo{volume}{450}},
  \bibinfo{pages}{857} (\bibinfo{year}{2007}).

\bibitem[{\citenamefont{Fushman et~al.}(2008)\citenamefont{Fushman, Englund,
  Faraon, Stoltz, Petroff, and Vuckovic}}]{FushmanSci08}
\bibinfo{author}{\bibfnamefont{I.}~\bibnamefont{Fushman}},
  \bibinfo{author}{\bibfnamefont{D.}~\bibnamefont{Englund}},
  \bibinfo{author}{\bibfnamefont{A.}~\bibnamefont{Faraon}},
  \bibinfo{author}{\bibfnamefont{N.}~\bibnamefont{Stoltz}},
  \bibinfo{author}{\bibfnamefont{P.}~\bibnamefont{Petroff}}, \bibnamefont{and}
  \bibinfo{author}{\bibfnamefont{J.}~\bibnamefont{Vuckovic}},
  \bibinfo{journal}{Science} \textbf{\bibinfo{volume}{320}},
  \bibinfo{pages}{769} (\bibinfo{year}{2008}).

\bibitem[{\citenamefont{Imamo\u{g}lu et~al.}(1999)\citenamefont{Imamo\u{g}lu,
  Awschalom, Burkard, DiVincenzo, Loss, Sherwin, and Small}}]{ImamogluPRL99}
\bibinfo{author}{\bibfnamefont{A.}~\bibnamefont{Imamo\u{g}lu}},
  \bibinfo{author}{\bibfnamefont{D.~D.} \bibnamefont{Awschalom}},
  \bibinfo{author}{\bibfnamefont{G.}~\bibnamefont{Burkard}},
  \bibinfo{author}{\bibfnamefont{D.~P.} \bibnamefont{DiVincenzo}},
  \bibinfo{author}{\bibfnamefont{D.}~\bibnamefont{Loss}},
  \bibinfo{author}{\bibfnamefont{M.}~\bibnamefont{Sherwin}}, \bibnamefont{and}
  \bibinfo{author}{\bibfnamefont{A.}~\bibnamefont{Small}},
  \bibinfo{journal}{\prl} \textbf{\bibinfo{volume}{83}}, \bibinfo{pages}{4204}
  (\bibinfo{year}{1999}).

\bibitem[{\citenamefont{Sridharan and Waks}(2008)}]{SridharanPRA08}
\bibinfo{author}{\bibfnamefont{D.}~\bibnamefont{Sridharan}} \bibnamefont{and}
  \bibinfo{author}{\bibfnamefont{E.}~\bibnamefont{Waks}},
  \bibinfo{journal}{Phys. Rev. A} \textbf{\bibinfo{volume}{78}},
  \bibinfo{pages}{052321} (\bibinfo{year}{2008}).

\bibitem[{\citenamefont{Bayer et~al.}(2002)\citenamefont{Bayer, Ortner, Stern,
  Kuther, Gorbunov, Forchel, Hawrylak, Fafard, Hinzer, Reinecke
  et~al.}}]{BayerPRB02}
\bibinfo{author}{\bibfnamefont{M.}~\bibnamefont{Bayer}},
  \bibinfo{author}{\bibfnamefont{G.}~\bibnamefont{Ortner}},
  \bibinfo{author}{\bibfnamefont{O.}~\bibnamefont{Stern}},
  \bibinfo{author}{\bibfnamefont{A.}~\bibnamefont{Kuther}},
  \bibinfo{author}{\bibfnamefont{A.~A.} \bibnamefont{Gorbunov}},
  \bibinfo{author}{\bibfnamefont{A.}~\bibnamefont{Forchel}},
  \bibinfo{author}{\bibfnamefont{P.}~\bibnamefont{Hawrylak}},
  \bibinfo{author}{\bibfnamefont{S.}~\bibnamefont{Fafard}},
  \bibinfo{author}{\bibfnamefont{K.}~\bibnamefont{Hinzer}},
  \bibinfo{author}{\bibfnamefont{T.~L.} \bibnamefont{Reinecke}},
  \bibnamefont{et~al.}, \bibinfo{journal}{Phys. Rev. B}
  \textbf{\bibinfo{volume}{65}}, \bibinfo{pages}{195315}
  (\bibinfo{year}{2002}).

\bibitem[{\citenamefont{Xu et~al.}(2008)\citenamefont{Xu, Sun, Berman, Steel,
  Bracker, Gammon, and Sham}}]{XuNPhys08}
\bibinfo{author}{\bibfnamefont{X.}~\bibnamefont{Xu}},
  \bibinfo{author}{\bibfnamefont{B.}~\bibnamefont{Sun}},
  \bibinfo{author}{\bibfnamefont{P.~R.} \bibnamefont{Berman}},
  \bibinfo{author}{\bibfnamefont{D.~G.} \bibnamefont{Steel}},
  \bibinfo{author}{\bibfnamefont{A.~S.} \bibnamefont{Bracker}},
  \bibinfo{author}{\bibfnamefont{D.}~\bibnamefont{Gammon}}, \bibnamefont{and}
  \bibinfo{author}{\bibfnamefont{L.~J.} \bibnamefont{Sham}},
  \bibinfo{journal}{Nature Physics} \textbf{\bibinfo{volume}{4}},
  \bibinfo{pages}{692} (\bibinfo{year}{2008}).

\bibitem[{\citenamefont{Press et~al.}(2008)\citenamefont{Press, Ladd, Zhang,
  and Yamamoto}}]{PressNat08}
\bibinfo{author}{\bibfnamefont{D.}~\bibnamefont{Press}},
  \bibinfo{author}{\bibfnamefont{T.~D.} \bibnamefont{Ladd}},
  \bibinfo{author}{\bibfnamefont{B.}~\bibnamefont{Zhang}}, \bibnamefont{and}
  \bibinfo{author}{\bibfnamefont{Y.}~\bibnamefont{Yamamoto}},
  \bibinfo{journal}{Nature (London)} \textbf{\bibinfo{volume}{456}},
  \bibinfo{pages}{218} (\bibinfo{year}{2008}).

\bibitem[{\citenamefont{Kiraz et~al.}(2004)\citenamefont{Kiraz, Atat\"ure, and
  Imamo\ifmmode~\breve{g}\else \u{g}\fi{}lu}}]{KirazPRA04}
\bibinfo{author}{\bibfnamefont{A.}~\bibnamefont{Kiraz}},
  \bibinfo{author}{\bibfnamefont{M.}~\bibnamefont{Atat\"ure}},
  \bibnamefont{and}
  \bibinfo{author}{\bibfnamefont{A.}~\bibnamefont{Imamo\ifmmode~\breve{g}\else
  \u{g}\fi{}lu}}, \bibinfo{journal}{Phys. Rev. A}
  \textbf{\bibinfo{volume}{69}}, \bibinfo{pages}{032305}
  (\bibinfo{year}{2004}).

\bibitem[{\citenamefont{Faraon et~al.}(2007)\citenamefont{Faraon, Englund,
  Fushman, Vu\v{c}kovi\'{c}, Stoltz, and Petroff}}]{FaraonAPL07}
\bibinfo{author}{\bibfnamefont{A.}~\bibnamefont{Faraon}},
  \bibinfo{author}{\bibfnamefont{D.}~\bibnamefont{Englund}},
  \bibinfo{author}{\bibfnamefont{I.}~\bibnamefont{Fushman}},
  \bibinfo{author}{\bibfnamefont{J.}~\bibnamefont{Vu\v{c}kovi\'{c}}},
  \bibinfo{author}{\bibfnamefont{N.}~\bibnamefont{Stoltz}}, \bibnamefont{and}
  \bibinfo{author}{\bibfnamefont{P.}~\bibnamefont{Petroff}},
  \bibinfo{journal}{\apl} \textbf{\bibinfo{volume}{90}},
  \bibinfo{pages}{213110} (\bibinfo{year}{2007}).

\bibitem[{\citenamefont{Mosor et~al.}(2005)\citenamefont{Mosor, Hendrickson,
  Richards, Sweet, Khitrova, Gibbs, Yoshie, Scherer, Shchekin, and
  Deppe}}]{MosorAPL05}
\bibinfo{author}{\bibfnamefont{S.}~\bibnamefont{Mosor}},
  \bibinfo{author}{\bibfnamefont{J.}~\bibnamefont{Hendrickson}},
  \bibinfo{author}{\bibfnamefont{B.~C.} \bibnamefont{Richards}},
  \bibinfo{author}{\bibfnamefont{J.}~\bibnamefont{Sweet}},
  \bibinfo{author}{\bibfnamefont{G.}~\bibnamefont{Khitrova}},
  \bibinfo{author}{\bibfnamefont{H.~M.} \bibnamefont{Gibbs}},
  \bibinfo{author}{\bibfnamefont{T.}~\bibnamefont{Yoshie}},
  \bibinfo{author}{\bibfnamefont{A.}~\bibnamefont{Scherer}},
  \bibinfo{author}{\bibfnamefont{O.~B.} \bibnamefont{Shchekin}},
  \bibnamefont{and} \bibinfo{author}{\bibfnamefont{D.~G.} \bibnamefont{Deppe}},
  \bibinfo{journal}{\apl} \textbf{\bibinfo{volume}{87}},
  \bibinfo{pages}{141105} (\bibinfo{year}{2005}).

\bibitem[{\citenamefont{{Strauf} et~al.}(2006)\citenamefont{{Strauf}, {Rakher},
  {Carmeli}, {Hennessy}, {Meier}, {Badolato}, {Dedood}, {Petroff}, {Hu},
  {Gwinn} et~al.}}]{StraufAPL06}
\bibinfo{author}{\bibfnamefont{S.}~\bibnamefont{{Strauf}}},
  \bibinfo{author}{\bibfnamefont{M.~T.} \bibnamefont{{Rakher}}},
  \bibinfo{author}{\bibfnamefont{I.}~\bibnamefont{{Carmeli}}},
  \bibinfo{author}{\bibfnamefont{K.}~\bibnamefont{{Hennessy}}},
  \bibinfo{author}{\bibfnamefont{C.}~\bibnamefont{{Meier}}},
  \bibinfo{author}{\bibfnamefont{A.}~\bibnamefont{{Badolato}}},
  \bibinfo{author}{\bibfnamefont{M.~J.~A.} \bibnamefont{{Dedood}}},
  \bibinfo{author}{\bibfnamefont{P.~M.} \bibnamefont{{Petroff}}},
  \bibinfo{author}{\bibfnamefont{E.~L.} \bibnamefont{{Hu}}},
  \bibinfo{author}{\bibfnamefont{E.~G.} \bibnamefont{{Gwinn}}},
  \bibnamefont{et~al.}, \bibinfo{journal}{\apl} \textbf{\bibinfo{volume}{88}},
  \bibinfo{pages}{043116} (\bibinfo{year}{2006}).

\bibitem[{\citenamefont{Laucht et~al.}(2009)\citenamefont{Laucht, Hofbauer,
  Hauke, Angele, Stobbe, Kaniber, Bohm, Lodahl, Amann, and
  Finley}}]{LauchtNJP09}
\bibinfo{author}{\bibfnamefont{A.}~\bibnamefont{Laucht}},
  \bibinfo{author}{\bibfnamefont{F.}~\bibnamefont{Hofbauer}},
  \bibinfo{author}{\bibfnamefont{N.}~\bibnamefont{Hauke}},
  \bibinfo{author}{\bibfnamefont{J.}~\bibnamefont{Angele}},
  \bibinfo{author}{\bibfnamefont{S.}~\bibnamefont{Stobbe}},
  \bibinfo{author}{\bibfnamefont{M.}~\bibnamefont{Kaniber}},
  \bibinfo{author}{\bibfnamefont{G.}~\bibnamefont{Bohm}},
  \bibinfo{author}{\bibfnamefont{P.}~\bibnamefont{Lodahl}},
  \bibinfo{author}{\bibfnamefont{M.~C.} \bibnamefont{Amann}}, \bibnamefont{and}
  \bibinfo{author}{\bibfnamefont{J.~J.} \bibnamefont{Finley}},
  \bibinfo{journal}{New Journal of Physics} \textbf{\bibinfo{volume}{11}},
  \bibinfo{pages}{023034} (\bibinfo{year}{2009}).

\bibitem[{\citenamefont{Kim et~al.}(2009)\citenamefont{Kim, Thon, Petroff, and
  Bouwmeester}}]{KimAPL09}
\bibinfo{author}{\bibfnamefont{H.}~\bibnamefont{Kim}},
  \bibinfo{author}{\bibfnamefont{S.~M.} \bibnamefont{Thon}},
  \bibinfo{author}{\bibfnamefont{P.~M.} \bibnamefont{Petroff}},
  \bibnamefont{and}
  \bibinfo{author}{\bibfnamefont{D.}~\bibnamefont{Bouwmeester}},
  \bibinfo{journal}{\apl} \textbf{\bibinfo{volume}{95}},
  \bibinfo{pages}{243107} (\bibinfo{year}{2009}).

\bibitem[{\citenamefont{Sridharan et~al.}(2010)\citenamefont{Sridharan, Waks,
  Solomon, and Fourkas}}]{SridharanAPL10}
\bibinfo{author}{\bibfnamefont{D.}~\bibnamefont{Sridharan}},
  \bibinfo{author}{\bibfnamefont{E.}~\bibnamefont{Waks}},
  \bibinfo{author}{\bibfnamefont{G.}~\bibnamefont{Solomon}}, \bibnamefont{and}
  \bibinfo{author}{\bibfnamefont{J.~T.} \bibnamefont{Fourkas}},
  \bibinfo{journal}{\apl} \textbf{\bibinfo{volume}{96}},
  \bibinfo{pages}{153303} (\bibinfo{year}{2010}).

\bibitem[{\citenamefont{Reitzenstein et~al.}(2009)\citenamefont{Reitzenstein,
  M\"unch, Franeck, Rahimi-Iman, L\"offler, H\"ofling, Worschech, and
  Forchel}}]{ReitzensteinPRL09}
\bibinfo{author}{\bibfnamefont{S.}~\bibnamefont{Reitzenstein}},
  \bibinfo{author}{\bibfnamefont{S.}~\bibnamefont{M\"unch}},
  \bibinfo{author}{\bibfnamefont{P.}~\bibnamefont{Franeck}},
  \bibinfo{author}{\bibfnamefont{A.}~\bibnamefont{Rahimi-Iman}},
  \bibinfo{author}{\bibfnamefont{A.}~\bibnamefont{L\"offler}},
  \bibinfo{author}{\bibfnamefont{S.}~\bibnamefont{H\"ofling}},
  \bibinfo{author}{\bibfnamefont{L.}~\bibnamefont{Worschech}},
  \bibnamefont{and} \bibinfo{author}{\bibfnamefont{A.}~\bibnamefont{Forchel}},
  \bibinfo{journal}{Phys. Rev. Lett.} \textbf{\bibinfo{volume}{103}},
  \bibinfo{pages}{127401} (\bibinfo{year}{2009}).

\bibitem[{\citenamefont{Akahane et~al.}(2005)\citenamefont{Akahane, Asano,
  Song, and Noda}}]{AkahaneOptExp05}
\bibinfo{author}{\bibfnamefont{Y.}~\bibnamefont{Akahane}},
  \bibinfo{author}{\bibfnamefont{T.}~\bibnamefont{Asano}},
  \bibinfo{author}{\bibfnamefont{B.-S.} \bibnamefont{Song}}, \bibnamefont{and}
  \bibinfo{author}{\bibfnamefont{S.}~\bibnamefont{Noda}},
  \bibinfo{journal}{Optics Express} \textbf{\bibinfo{volume}{13}},
  \bibinfo{pages}{1202} (\bibinfo{year}{2005}).

\bibitem[{\citenamefont{Walck and Reinecke}(1998)}]{WalckPRB98}
\bibinfo{author}{\bibfnamefont{S.~N.} \bibnamefont{Walck}} \bibnamefont{and}
  \bibinfo{author}{\bibfnamefont{T.~L.} \bibnamefont{Reinecke}},
  \bibinfo{journal}{Phys. Rev. B} \textbf{\bibinfo{volume}{57}},
  \bibinfo{pages}{9088} (\bibinfo{year}{1998}).

\bibitem[{\citenamefont{Krizhanovskii et~al.}(2005)\citenamefont{Krizhanovskii,
  Ebbens, Tartakovskii, Pulizzi, Wright, Skolnick, and
  Hopkinson}}]{KrizhanovskiiPRB05}
\bibinfo{author}{\bibfnamefont{D.~N.} \bibnamefont{Krizhanovskii}},
  \bibinfo{author}{\bibfnamefont{A.}~\bibnamefont{Ebbens}},
  \bibinfo{author}{\bibfnamefont{A.~I.} \bibnamefont{Tartakovskii}},
  \bibinfo{author}{\bibfnamefont{F.}~\bibnamefont{Pulizzi}},
  \bibinfo{author}{\bibfnamefont{T.}~\bibnamefont{Wright}},
  \bibinfo{author}{\bibfnamefont{M.~S.} \bibnamefont{Skolnick}},
  \bibnamefont{and}
  \bibinfo{author}{\bibfnamefont{M.}~\bibnamefont{Hopkinson}},
  \bibinfo{journal}{Phys. Rev. B} \textbf{\bibinfo{volume}{72}},
  \bibinfo{pages}{161312} (\bibinfo{year}{2005}).

\bibitem[{\citenamefont{Muller et~al.}(2009)\citenamefont{Muller, Fang, Lawall,
  and Solomon}}]{MullerPRL09}
\bibinfo{author}{\bibfnamefont{A.}~\bibnamefont{Muller}},
  \bibinfo{author}{\bibfnamefont{W.}~\bibnamefont{Fang}},
  \bibinfo{author}{\bibfnamefont{J.}~\bibnamefont{Lawall}}, \bibnamefont{and}
  \bibinfo{author}{\bibfnamefont{G.~S.} \bibnamefont{Solomon}},
  \bibinfo{journal}{Phys. Rev. Lett.} \textbf{\bibinfo{volume}{103}},
  \bibinfo{pages}{217402} (\bibinfo{year}{2009}).

\bibitem[{HKi()}]{HKim}
\bibinfo{note}{Oscillator strength of typical InAs QDs is on the order of 10,
  which corresponds to a lateral extension of $\sim 7$~nm of electron and hole
  wave function. The magnetic confinement at 7~T is about 10~nm.}

\end{thebibliography}
\end{document}